\documentclass[]{spie}  

\usepackage{amsmath,amsfonts,amssymb}
\usepackage[numbers,sort&compress]{natbib}
\usepackage{graphicx}
\usepackage{hyperref} 
\usepackage{etoolbox}
\apptocmd{\thebibliography}{\setlength{\itemsep}{-2pt}}{}{}

\usepackage[justification=centering]{caption}

\title{Domain Shift in Computer Vision models for MRI data analysis: An Overview}

\author[a]{Ekaterina Kondrateva*}
\author[a]{Marina Pominova}
\author[a]{Elena Popova}
\author[a]{Maxim Sharaev}
\author[a]{Alexander Bernstein}
\author[a]{Evgeny Burnaev}
\affil[a]{Skolkovo Institute of Science and Technology}

\begin{document} 
\maketitle

\begin{abstract}
Machine learning and computer vision methods are showing good performance in medical imagery analysis. Yet only a few applications are now in clinical use and one of the reasons for that is poor transferability of the models to data from different sources or acquisition domains. Development of new methods and algorithms for the transfer of training and adaptation of the domain in multi-modal medical imaging data is crucial for the development of accurate models and their use in clinics. In present work, we overview methods used to tackle the domain shift problem in machine learning and computer vision. The algorithms discussed in this survey include advanced data processing, model architecture enhancing and featured training, as well as predicting in domain invariant latent space. The application of the autoencoding neural networks and their domain-invariant variations are heavily discussed in a survey. We observe the latest methods applied to the magnetic resonance imaging (MRI) data analysis and conclude on their performance as well as propose directions for further research.
\end{abstract}


\section{INTRODUCTION}
\label{sec:intro}  

Image recognition systems and, in particular, machine learning methods have recently found many applications in the field of medical diagnostics and image processing \cite{bzdok2017inference}. 
Conventionally,  medical imaging data and MRI, in particular, are preprocessed to prepare for further analysis or to obtain lower-dimensional representation. In the first case, the data can be standardized or normalized and thus brought to a common form. In the second case, data preprocessing can be done for the extraction of anatomically or physiologically meaningful features from images. These features can be considered as potential biomarkers and used for diagnostics and localization of pathology.

Examples of such features for diffusion tensor imaging (DTI) are the values of fractional anisotropy (FA) and diffusivity (MD) of the signal in certain regions of the brain. Other examples of informative features are the values of functional connectivity between brain regions for functional MRI (fMRI) and morphometric characteristics for structural MRI. Machine learning models can be trained on the extracted characteristics, and most important or higher scored features can be regarded as potential biomarkers \cite{peng2017multilevel}.

However, deep learning approaches, especially convolutional networks (CNN), have proven to be more accurate in many applications \cite{wang2019dilated} because they use full-size data without losing information during preprocessing. Existing CNN models make it possible to analyze medical images, both in 2D and 3D \cite{ueda2019age}, which, in case of neuroimaging data, take into account their spatial structure and achieve a higher prediction accuracy.These computer vision methods are being successfully used for the ishemic stroke segmentation \cite{liu2019deep}, diagnosis of Alzheimer's disease \cite{wen2020convolutional, hon2017towards}, epilepsy \cite{Pominova2019Conv}, depressive disorders \cite{sharaev2018mri}, autism \cite{mellema2019multiple} and others.

Nevertheless only a few of the proposed methods have found their clinical application. For example, researchers from DeepMind (Google) presented the results of a clinical trial of a system for early detection of breast cancer based on computed tomography (CT) images \cite{mckinney2020international}, which showed a false positive rate of 5.7\% and 1.2\% (in the US and UK) and target missed rates of 9.4\% and 2.7\% respectively. This deep learning system was 11.5\% more accurate than the average radiologist in terms of AUC (Area Under The Curve) - ROC (Receiver Operating Characteristics) score. More importantly, the authors compared the predictive power of a system trained on data from the United Kingdom with the accuracy of a system trained on data from the United States.

The aforementioned study is one of the first large-scale international clinical trials of artificial intelligence-based decision support systems. At the same time, many of the smaller studies have already become commercial solutions for the problems of ophthalmology, fluorography, dentistry. Such technologies are used, for example, in the services "Botkin AI"\footnote{access https://botkin.ai/}, "Third Opinion"\footnote{access, https://thirdopinion.ai/}. However, these decisions are based on the restricted samples, with a more fixed range of modalities and scanning protocols.  This can be sufficient for the use in the particular hospital yet their use on the international scale can be questioned. 
For neuroimaging and, in particular, MRI data analysis, the development of domain-agnostic models is especially important, because MRI study is costly and not widely spread, and the data implies complicated analysis and often in 3D. Thus at the moment, there are no such solutions and large-scale studies for neuroimaging data, and the overwhelming number of published works are based on the analysis of only a (mostly) few tens or hundreds of images. The models trained on small samples of high-dimensional data tend to lose predictive ability when tested on similar data from another domain (scanning protocol), even if the same preprocessing pipeline \cite{Sharaev2018a} was used and is brought to a standard form. Thus the construction of neuroimaging domain-agnostic models an actual task.

\section{DOMAIN SHIFT PROBLEM IN DATA ANALYSIS}

In data analysis, if the training data represent an unbiased sample of the general distribution, then the constructed decision rule can be well generalized to new data and thus guarantee an accurate prediction for the new samples. However, if the training data is not representative, we can expect a difference between the distributions of the future test data and the original training sample. Standard ML models cannot cope with changes in the distribution of data between training stages and thus lose their accuracy when tested on unseen data. Domain adaptation, transfer of training, and preliminary training of models from one sample to another are areas of machine learning that are aimed at solving this domain shift problem.

All domain adaptation methods can be roughly divided into three categories: data transformation (normalization), the transformation of a model or training conditions, and search in the space of characteristics independent to the domain.

\begin{figure} [ht]
\begin{center}
\begin{tabular}{c} 
\includegraphics[height=8cm]{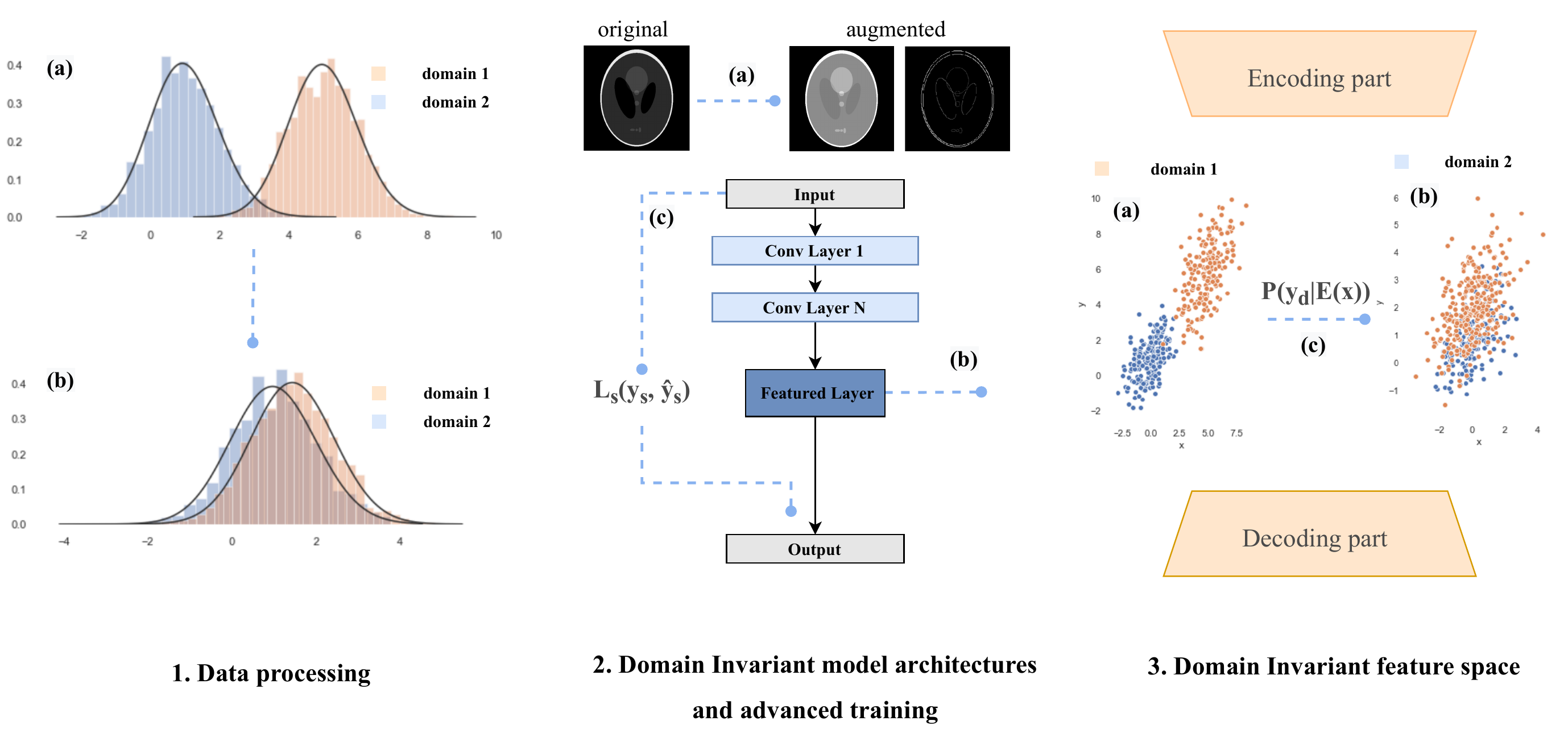}
\end{tabular}

\caption[example] 
{ 
\label{fig:main} 
 Three approaches to tackle the domain shift problem in machine learning: 1. Domain shift in data feature distributions from different sources before (a) and after (b) the correction or domain transfer; 
 2. Data augmentation for training (a); Featured layers, stable for small data distortions (b);  Adversarial loss functions, controlling learned patterns (c) and parallel learning streams. Where $L_s$ - additional loss function on specific target feature $y_s$, like shape patterns or edges; 
 3. Domain shift in data latent representations from different sources before (a) and after (b) domain correction. Domain discriminator loss function (c) on data $x$ and domain target $y_d$.}
\end{center}
\end{figure}

This section contains an overview of transfer learning, data preprocessing techniques used to tackle the domain shift problem. These methods could be divided into three subcategories represented on Figure \ref{fig:main} \textbf{1.(a)} and \textbf{1.(b)}.

\subsection{Data processing}
Medical imagery data sets are typically preprocessed for further analysis (for example, to reduce noise) or to obtain significant low-dimensional features.
Data preprocessing can include methods for standardizing the distribution of individual image intensities, for example,  Z-normalization \cite{bologna2019assessment}, or matching normalized intensities of all data in data-sets from different sources with Histogram Matching \cite{Nyul1999} or deep learning transferring methods.

\subsubsection{Z-normalisation, Histogram matching}
Usually, machine learning algorithms are aimed at finding regularity in data comparing features of different data points. Wherein, a difficulty arises when the investigated functions have very different scales \cite{nyul2000new}. Conventionally all images are standardized in size and pixels intensity distributions individually with Intensity re-scaling and Histogram transformation methods. Further preprocessing can include Bias Field correction and image denoizing \cite{mukherjee2013efficient}.

Z-normalization represents a data point alignment procedure so that all considered features become equally important. Z-normalization was first proposed in \cite{goldin1995similarity}. As a result of applying this method, all elements of the input vector are converted into an output vector in such a way that its average value becomes approximately equal to 0, and the standard deviation becomes close to 1. Note that such normalization cleans data well from outliers, while normalized data does not have the same scale.

Histogram matching or histogram standardization is used in image processing to match all intensity distributions on the sample to a specified histogram \cite{ekstrom2012digital}. A common application of this method is to match the images from two sensors with slightly different responses, or from a sensor whose response changes over time. 

Another approach has been proposed to match distributions on features, retrieved with Principal Component Analysis (PCA) from full-sized 
fMRI data. The based on \cite{massey1951kolmogorov} method yields the winning solution in Trends\footnote{https://www.kaggle.com/c/trends-assessment-prediction} competition, there the goal was to predict the age and physiological characteristics of the subject irrespectively to the MRI data domain. 

To involve the matching of histograms for MRI normalization was proposed by \cite{Nyul1999} and refined by \cite{shah2011evaluating} and \cite{nyul2000new}. A histogram matching method was used for correcting the variations in scanner sensitivity in \cite{collewet2004influence},  the efficiency of the analysis of the derivatives of even order on the MRI histogram was shown in \cite{wang1998correction}. Some histogram matching algorithms are demonstrated in \cite{jager2008nonrigid} for comparing the histogram of the input image with the histogram of the reference image by minimizing information-oriented criteria. In the case of noise, the histogram match can depend on the number of bins \cite{roy2013patch}. 


\subsubsection{Neural Transfer Architectures}
\label{sec:title1}

One of the common tasks in the field of graphics and vision is the image-to-image translation which aims to create a newly generated version of a given image with some specified changes. The most common difficulty in using this method is preparing the required datasets, which must be large and often difficult or almost impossible to prepare. The cycle generative adversarial network (CycleGAN) is a technique with GAN architecture \cite{ krizhevsky2012proceedings}, which overcomes this difficulty using unrelated images from the source and target domains to train the models in an unsupervised manner. The algorithms could be found here   \footnote{https://www.tensorflow.org/tutorials/generative/cyclegan.}. 

In \cite{lei2019mri} a novel learning‐based approach was developed and validated to create CT images from MR images based on the CycleGAN model to find the relationship between the CT and MRI. In \cite{gong2019dlow} authors built a domain flow generation (DLOW) model to connect two different domains by creating a continuous sequence of intermediate domains passing from one domain to another. This model can transfer source images into different styles in the intermediate domains generating new image styles that are not visible in the training data.

\subsection{Domain Invariant Model Architectures and Advanced Training}
Another approach is the development domain-invariant and generalizable architectures and the use of advanced training methods. These include: data augmentation and model transfer for training as showed on Figure \ref{fig:main}, \textbf{2.(a)}; the use of specialised featured layers as adaptive receptive fields or topological feature extraction, see Figure \ref{fig:main}, \textbf{2.(b)}; or the use of architectures with additional training streams or combined loss functions see Figure \ref{fig:main}, \textbf{2.(c)}. All the mentioned approaches facilitate the learning of more conservative patterns or make the model more stable to data distortions.

\subsubsection{Data Augmentation}
Data augmentation is one of the most widely used advanced training methods in computer vision. The augmented dataset represents a more complete set of possible data samples, increasing model convergence while training and minimizing the distance between the domains. Data Augmentation approaches include basic manipulations with images like rotations, small deformations, and others, as well as deep learning approaches with neural style transfer and GANs \cite{shorten2019survey}. These latter methods were highlighted in the previous section yet in the context of translation data from a different domain to one from the training sample. In the meaning of data augmentation, these methods are used to train more stable and generalizable models with no further manipulations on future data. Data augmentations can be automated as in Smart Augmentation \cite{lemley2017smart} or introduce rationally expected distortions in test data. In case of MRI  - magnetic field distortions, noise and contrast could be augmented to increase model generalizability on data from other domains and build so-called "contrast-agnostic" models \cite{billot2020learning}.

It has been shown that CNNs in general are attentive to texture more than to global object shape\cite{baker2018deep}, so shapes are more conservative and informative features of am image. In neuroimaging data, different image contrasts and small patterns (as a result of, for example, field inhomogeneities) are to be more easily learned than particular organ shapes. Data augmentation helps to tackle this problem. Yet a different approach was shown in paper \cite{fetit2020training} where authors build semantic segmentation models on texture maps, retrieved from brain T2 scans. The results on 10 pixels texture map (0.981 DICE score) were comparable to ones from segmentation in grayscale images (0.990 DICE score) suggesting a more important role to image texture in model prediction. 

\subsubsection{Advanced training}
Apart from data augmentation, special training methods can enhance model generalizability. The most common approach is the use of combined loss functions in parallel learning streams. Then in a parallel stream, the model is forced to learn more conservative patterns.
An example of such a loss function is an Edge-aware network architecture \cite{guan2018edge}, which can be utilized to facilitate the learning of shape patterns in medical imaging. As in \cite{sun2020saunet} authors show the effectiveness of such an approach on MRI segmentation, increasing the segmentation of the right ventricle from 91.40 to 93.11 in DICE score.

Minimization of distribution differences is a basic approach to domain adaptation for imagery data analysis in classification, detection, and segmentation tasks. And this idea could be realized as an extra loss function, incorporated in the model architecture.  In \cite{pichler2019direct} a simple direct distribution matching approach was proposed for unsupervised domain adaptation in the context of semantic segmentation of medical images. Unlike all adversarial approaches in this domain, the proposed method aligns data kernel density distributions from domains in the label space. Thus optimizing the KL-divergence between two probability distributions from source $s$  and target image $s'$, as $D_KL(s,s')$, the probability vectors from model soft-max outputs obtained for each pixel. This output space conveys much richer local and global information, which results in better-adapted models. Authors have demonstrated that directly matching output distributions they were close to supervised AdaptSegNet\cite{Tsai_adaptseg_2018} in cross-domain MRI segmentation on MRBrainS dataset - 0.76 compared to 0.56 (DICE score).

\subsubsection{Transfer learning}
Transfer learning (TL) is a machine learning method that allows using the experience and learned patterns from one domain to solve another, similar problem. The weights and biases that a model uses to detect features in one domain often work well for detecting features in another domain if the domains are similar. So, the model is first trained on a large available amount of data, then tuned on a small target dataset.  

There is a number of successful applications of transfer learning to MRI analysis in the following works: \cite{yuan2019prostate},  \cite{chen2019transfer}, \cite{lu2019pathological}. For example, in one study the authors explore the effect of model "forgetting" while shifting the domains, showing that MRI brain parcellation is more accurate and better initialized with model transfer, and achieve 0.58 DICE training on 4\% of data, compared to no model transfer.


\subsubsection{Featured layers}
Featured layers on model architecture mostly represent adaptive receptive fields. Thus the convolution could be more stable to local data distortions. The basic approach used here is dilated convolutions, thus receptive "window" of the kernel became larger without excessive neurons in the networks, allowing to learn larger geometrical patterns \cite{moeskops2017adversarial}. 
Transformable convolutions \cite{xiao2018transformable}, containing the prior information of patterns are mostly used in text classification. Yet, deformable convolutions \cite{pominova20193d} were shown to be more robust on MRI imaging, increasing the accuracy of the classification from 0.788 to 0.823 AUC in 3D MRI Schizophrenia classification.

Another common featured layers are not yet adopted in MRI analysis in topologically invariant architectures \cite{zhang2018machine}. These featured layers show their applicability in docking and time series analysis \cite{liu2016applying}. The methods exhibit high potential for retrieving conservative and domain-invariant features as well as the use of persistent homology, or image representation in graph notation.

\subsection{Domain Invariant feature space}

The third approach in the domain shift problem is the domain-adaptation as it commonly called. Conventional domain adaptation methods cover auto-encoding networks with latent space data transformation and modelling in domain-invariant feature space \ref{fig:main}, \textbf{3.(c)}.

\subsubsection{Domain-Adversarial Training of Neural Network}

Most of the methods in the third group employ the adversarial learning to force the model to learn domain-independent features. Ganin et al. \cite{ganin2016domain} were one of the first to suggest introducing an additional network - a domain discriminator - to transfer a pre-tained model to a new domain. In Domain-Adversarial Training of Neural Networks (DATNN) the model for the main problem is trained only on a sample from “source domain”. At the same time, the discriminator takes features extracted by the model from both target and source domain data and tries to distinguish which domain these features come from.
As a result, the model learns to extract the features that confuse the discriminator most and complicate distinguishing between domains.


Originally, this technique allows domain adaptation in a semi-supervised manner when the labels for data from the target domain are not available. However, it can be also employed to train a domain-independent model on a wholly labeled multi-domain sample.


\subsubsection{Autoencoders, Fader Networks, DIVA}


The disadvantage of the previous method is the difficulty of stabilizing an adversarial training of the discriminator and the predictive model for the main problem. In such cases, the task can be simplified by dividing it into two sequential steps - first, extracting domain-invariant features with an autoencoder and second, using them to train a separate predictive model. 

Conventionally, an autoencoder consists of two parts - an encoder, which compresses the input image into a lower-dimensional vector representation, and a decoder, which receives this latent representation and tries to recover an original image. To make the encoder learn domain-independent latent representations, it can be trained in a way proposed by Lample et al. in \cite{lample2017fader}. Similar to the domain adaptation approaches, the authors introduce an additional discriminator network that operates in the latent space of autoencoder and predicts the value of some attribute variable from the latent representation. Since the discriminator is optimized adversarially with the encoder, the latter has to filter out the information related to the attribute. At the same time, the decoder receives attribute value as an additional input along to be able to correctly reconstruct an input image.

The whole model called Fader Network consists of the encoder, the decoder, and the discriminator and allows generating different versions of input image conditioned on the chosen attribute values, while the latent representation remains independent and does not contain any attribute-related information. However, due to the latter property, it can be employed for extracting domain-invariant features from multi-domain data, if the domain label is considered as the attribute variable.



Another variation of this approach is a DIVA (Domain Invariant Variational Autoencoder) model, proposed by \cite{ilse2019diva}. While Fader Network simply attempts to eliminate domain-related information from the entire latent vector $z$, DIVA partitions the latent space into three disjoint components - $z_d$, $z_x$, and $z_y$. The former is intended to store the part of data variability related to the domain, the latter - to the target variable, and $z_x$ encodes residual variability, independent from both target and domain. The whole model consists of three separate encoders that extract from the input image three parts of latent representation $z_d$, $z_x$, and $z_y$, the decoder transforms the entire vector $z = (z_d, z_x, z_y)$ back to the original image, and two auxiliary models try to predict values of the domain label $d$ and the target variable $y$ from $z_d$ and $z_y$ respectively. Note that in this case, training does not have an explicit adversarial component. However, DIVA is based on a variational autoencoder rather than a regular one, and the regularization of the latent space with a prior distribution limits the capacity of each vector $z$ component forcing it to optimally distribute information between $z_d$, $z_x$ and $z_y$.

\section{CONCLUSION}

In this work, we performed a comprehensive overview of the domain shift elimination methods in data analysis and their application to medical images and MRI in particular. We propose the classification of all the methods from the perspective of their action: data correction, model and training enhancement, and domain-invariant features search.

The three different methods can be potentially used in a row to facilitate the domain adaptation, yet these studies were not reported. On the other hand, data normalization techniques like Z-normalization and Histogram Matching are conventionally used in most of the MRI-applied studies. The existing works on deep learning style transfer in CT and MRI, on the contrary, appear to be unstable and data greedy methods.

Building domain-stable deep learning models and architectures is a matter of research in general computer vision. The most widely used approach here is data augmentation before training, this is also conventional for most MRI-segmentation models. Other approaches include adaptive receptive fields (deformable convolutions) and other featured layers as TDA, as well as specialized loss functions for shape attention.

The most rapidly developing approach in the domain shift problem is the domain-invariant autoencoding models, which were lately proposed and used from medical segmentation, as Fader network and DIVA and domain-agnostic model architectures.

\section{Acknowledgements} 
  The reported study was funded by RFBR according to the research project \textbf{20-37-90149}.



 

\bibliography{report} 

\bibliographystyle{spiebib} 

\end{document}